\documentclass[%
 aip,
 amsmath,amssymb,
 reprint,%
]{revtex4-1}

\usepackage{graphicx}
\usepackage{dcolumn}
\usepackage{bm}

\usepackage[utf8]{inputenc}
\usepackage[T1]{fontenc}
\usepackage{mathptmx}

\newcommand{\matr}[1]{{{\mathbf{#1}}}}    
\renewcommand{\vec}[1]{{\mathbf{#1}}}
\newcommand{\m}[1]{\begin{pmatrix}#1\end{pmatrix}}

\begin{document}

\preprint{AIP/123-QED}

\title[Synchronization in three-layer networks with a hub]{Synchronization scenarios in three-layer networks with a hub}

\author{Jakub Sawicki}
\email{zergon@gmx.net}
 \affiliation{Potsdam Institute for Climate Impact Research, Telegrafenberg A 31, 14473 Potsdam, Germany}

\author{Julia M. Koulen}%
 \affiliation{Potsdam Institute for Climate Impact Research, Telegrafenberg A 31, 14473 Potsdam, Germany}

\author{Eckehard Sch\"oll}
 \affiliation{Potsdam Institute for Climate Impact Research, Telegrafenberg A 31, 14473 Potsdam, Germany}
\affiliation{%
Institut f{\"u}r Theoretische Physik, Technische Universit\"at Berlin, Hardenbergstra\ss{}e 36, 10623 Berlin, Germany
}%

\affiliation{Bernstein Center for Computational Neuroscience Berlin, Humboldt-Universit{\"a}t, Philippstra{\ss}e 13, 10115 Berlin, Germany}

\date{\today}

\begin{abstract}
We study various relay synchronization scenarios in a three-layer network, where the middle (relay) layer is a single node, i.e. a hub, The two remote layers consist of non-locally coupled rings of FitzHugh-Nagumo oscillators modelling neuronal dynamics. All nodes of the remote layers are connected to the hub. The role of the hub and its importance for the existence of chimera states is investigated in dependence on the inter-layer coupling strength and inter-layer time delay. Tongue-like regions in the parameter plane exhibiting double chimeras, i.e., chimera states in the remote layers whose coherent cores are synchronized with each other, and salt-and-pepper states are found. At very low intra-layer coupling strength, 
when chimera states do not exist in single layers, these may be induced by the hub. Also the influence of dilution of links between the remote layers and the hub upon the dynamics is investigated. The greatest effect of 
dilution is observed when links to the coherent domain of the chimeras are removed.

\end{abstract}

\maketitle

\begin{quotation}

Relay synchronization plays an important role for neuronal dynamics. Especially, in the human brain structures exist where two different regions of the brain are connected to each other via a much smaller region, a hub. 
Physiological evidence shows various synchronization phenomena which can be related to the underlying network structure and have a big influence on the brain activity. In such relay networks the communication speed is affected by the distance between different regions and therefore information needs time to reach a remote region. It appears that the time delay as 
well as the number and strength of connections between the remote parts of the brain are relevant for the occurrence of synchronization patterns. These issues are the main subject of this study.

\end{quotation}

\section{\label{sec:Introduction}Introduction}

Many intriguing phenomena in nature can be described by means of nonlinear dynamics. The theory of such dynamical systems is a wide and independent field of scientific research~\footnote{We dedicate this paper to the memory of Vadim S.\,Anishchenko who has done pioneering work in this field}, including different aspects of synchronization and chaos control as well as the methods of reconstruction of attractors and dynamical systems from experimental time series~\cite{ANI95,ANI07,ANI14}. Especially, synchronization phenomena are ubiquitous in living systems and synchronization in networks of oscillators is of great current interest~\cite{PIK01,BOC18}. Recently, special attention has been paid to partial synchronization patterns~\cite{SCH16b,SCH20,SCH20b,SAW20,ZAK20,SCH20}, e.g., chimera states where incoherent and coherent oscillations occur in spatially coexisting domains~\cite{KUR02a,ABR04}. The surprising aspect of this phenomenon is that these states were detected in systems of identical oscillators coupled in a symmetric ring topology with a symmetric interaction function, and they coexist with a stable completely synchronized state. The last decade has seen an increasing interest in phase and amplitude chimeras both in time-continuous systems~\cite{PAN15,SEM16,ZAK17a,SHE17d,SHE18b,SHE20a,SHE21a,SHE21b,SHE21c} and in time-discrete maps~\cite{OME11,SEM15a,BOG16,BOG16a,VAD16,SEM17,SHE17,STR17b,SHE17c,BUK17,RYB17,SEM18a,BUK18,RYB18,BUK19,WIN19,RYB19,RYB19a,RYB20}. It has been shown that they are not limited to phase oscillators, but can be found for a large variety of different dynamics including neural systems~\cite{OME13,OME15,MAJ18a,CHO18,RAM19,GER20,SAW21a}. 

An important aspect in network science are multilayer networks, where recent research has opened up new facets, providing a description of systems 
interconnected through different types of links. One class of interactions are within the layers, and additionally other types of interactions occur between the network nodes from different layers~\cite{BOC14,DE13,DE15,KIV14}. A prominent example for such structures are neuronal networks, or 
social networks which can be described as groups of people with different 
patterns of contacts or interactions between them~\cite{GIR02}. Other relevant applications are communication, supply, and transportation networks, for instance power grids, subway networks, or airtraffic networks~\cite{CAR13d}. Moreover, multilayer networks are also known to generate and stabilize diverse partial synchronization patterns in adaptive networks~\cite{BER20,BER21}, where the connectivity changes in time~\cite{BER21c}.

A fascinating phenomenon in networks with multiplex topology is relay (or 
remote) synchronization between layers which are not directly connected, and interact via an intermediate (relay) layer~\cite{LEY18}. Remote synchronization, a regime where pairs of nodes synchronize despite their large 
distances on the network graph, has been shown to depend on the network symmetries~\cite{BER12,NIC13,GAM13,ZHA17,ZHA17a,ZHA20}. In neuroscience various scenarios have been uncovered where specific brain areas act as a functional relay between other brain regions, having a strong influence on 
signal propagation, brain functionality, and dysfunctions~\cite{ROE97,SOT06}. For instance, the relay cells of the thalamus serve both as the primary relay of sensory information from the periphery to the cortex and as an interactive hub of communication between cortical areas~\cite{SHE16b,RHO05,GUI02,GOL10a}. They enable visual processing~\cite{WAN11g}, and rapid coordination of spatially segregated cortical computations important for cognitive flexibility, cognitive control and its perturbation in disease states~\cite{HAL17}. Parahippocampal regions can be considered as relay 
stations, which actively gate impulse traffic between neocortex and hippocampus, with strong implications for the propagation of neural activity~\cite{CUR04}. The hippocampus also acts as a relay in the cortico-cortical 
theta synchronization~\cite{FIS06,GOL11a}; signal transmission between cortical and subcortical brain regions is involved in a wide range of brain 
functions~\cite{PRA13}. Especially partial relay synchronization plays an 
important role in experiments with mice~\cite{GOL11a}, where just a part of the hippocampal relay exhibits phase-lag synchronization with the two cortical regions, which between themselves exhibit partial zero-lag synchronization.

Going towards more realistic models, time-delay plays a significant role in the modeling of the dynamics of complex networks~\cite{SAW20}. In brain networks, the communication speed is affected by the distance between regions and therefore a stimulation applied to one region needs time to reach a different region. Thus, the introduction of time delay between the layers in a network is not only a factor towards more realistic modeling, 
but entering into the system delay time can also work as a powerful parameter to control dynamical patterns. 

A simple realization of a system which exhibits relay synchronization is a triplex network where a relay layer in the middle acts as a transmitter 
between the two outer layers. Recently the notion of relay synchronization has been extended from completely synchronized states to partial synchronization patterns in the individual layers of a three-layer multiplex network. It has been shown that the three-layer structure of the network allows for (partial) synchronization of chimera states in the outer layers via the relay layer~\cite{SAW18c,SAW18,SAW19a,WIN19,DRA20}. In this work we consider the special configuration where the relay layer is reduced to 
a single node, i.e., a hub. The dynamics on each node is given by the paradigmatic FitzHugh-Nagumo model of neuronal dynamics. In Sect.\,II we introduce the model. In Sect.\,III we investigate how the partial synchronization of chimera states depends upon the inter-layer coupling delay and strength. In Sect.\,IV we show that chimera states which do not exist in single layers can be induced by the hub. In Sect.\,V we analyze the influence of network topology, specifically we study how a dilution of the inter-layer connections between the outer layers and the hub plays a role for the 
chimera states in the outer layers. Sect.\,VI gives a conclusion.

\section{\label{sec:Model}Model}

\begin{figure}
\centering
\includegraphics[width=0.5\linewidth]{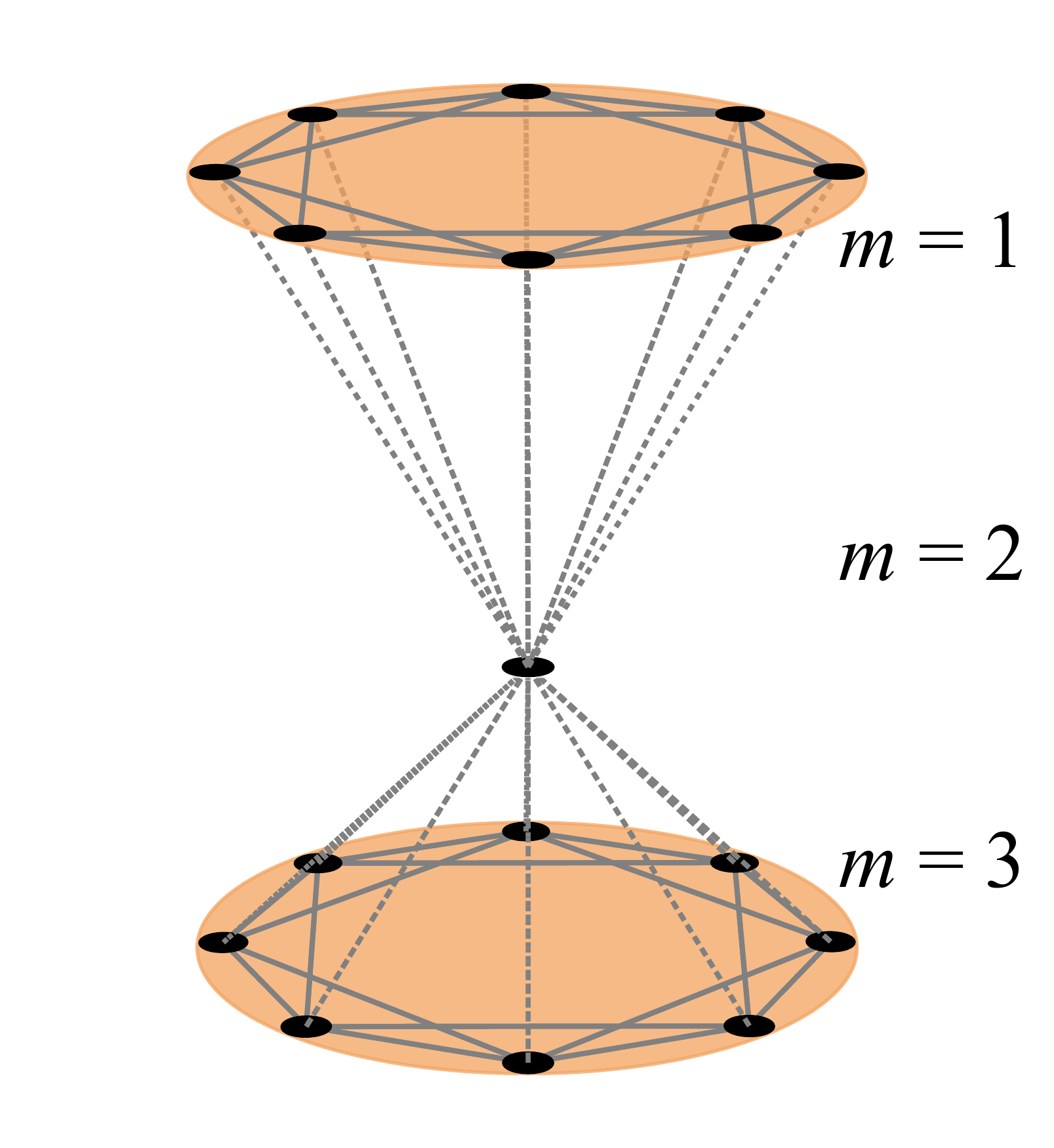}
\caption{Illustration of the three-layer network with a hub. All layers $m=1,2,3$ consist of identical FitzHugh-Nagumo oscillators, where each of the outer layers $m=1,3$ includes 500 nodes arranged in a non-local ring topology and the relay layer $m=2$ consists of a single node. The intra-layer coupling within the outer layers is instantaneous and is represented by solid lines, while the inter-layer coupling between the outer layers and the hub is time-delayed and is depicted by dotted lines.}
\label{fig:mymodel}
\end{figure}

We consider a three-layer network, in which the outer layers consist of a 
ring of $N$ identical FitzHugh-Nagumo (FHN) oscillators~\cite{FIT61,NAG62,BAS18} with non-local (intra-layer) coupling of coupling range $R^m$ and 
the middle layer consists of an individual node, i.e., a hub. An illustration of the considered model is presented in Fig.\,\ref{fig:mymodel}. Layers $m=1$ and $3$ (orange) are coupled through the intermediate layer $2$ (hub), so that the middle layer acts as a relay between the two outer layers, but there is no inter-layer coupling between layers $1$ and $3$. The dynamical equations for the outer layers ($m=1,3$) are given by
\begin{align}
\vec{\dot{x}}_i^{m}(t) = \vec{F}(\vec{x}_i^{m}(t)) &+ \frac{\sigma^m}{2R^m} \sum^{i+R^m}_{j=i-R^m}\matr{H}[\vec{x}_j^{m}(t)-\vec{x}_i^{m}(t)] + \nonumber \\
&+ \sum^{3}_{l=1} \sigma^{ml} \matr{G}[\vec{x}^2(t-\tau)-\vec{x}_i^{m}(t)],
\label{eq:my_equation_outer}
\end{align}
and for the hub ($m=2$) which acts as an active relay by
\begin{align}
\vec{\dot{x}}^{2}(t) = \vec{F}(\vec{x}^{2}(t)) + \sum^{3}_{l=1} \sum^{N}_{i=1} \frac{\sigma^{2l}}{N} \matr{G}[\vec{x}_i^{l}(t-\tau)-\vec{x}^2(t)],
\label{eq:my_equation_relay}
\end{align}
where $\vec{x}_i^m =(u, v)^T\in \mathbb{R}^2$, $m \in \{1,2,3\}$, $i \in \{1,...,N\}$ with all indices modulo $N$, denotes the set of activator ($u$) and inhibitor ($v$) variables, $\vec{x}_i^2 \equiv\vec{x}^2$ for the hub, and the dynamics of each individual oscillator is governed by 
\begin{eqnarray}
\label{eq:localdyn}
\vec{F}(\vec{x})=
\left(\!
\begin{array}{*{1}{c}}
\varepsilon^{-1}(u-\frac{u^3}{3}-v)\\
u + a
\end{array}
\!\right),
\end{eqnarray}
where $\varepsilon > 0$ describes the time scale separation between fast activator and slow inhibitor, fixed at $\varepsilon = 0.05$ throughout this work. Depending on the threshold parameter $a$, the single FHN elements exhibit either oscillatory ($|a|<1$) or excitable ($|a|>1$) behavior. 
Here we choose the oscillatory regime ($a=0.5$). The parameter~$\sigma^m$ denotes the intra-layer coupling strength ($\sigma_{\text{intra}}$), while $\sigma^{ml}$ is the inter-layer coupling strength ($\sigma_{\text{inter}}$). We use time delay $\tau$ only in the inter-layer coupling, since in real-world systems the transfer of information between two different 
layers is often slower than within one layer. In order to ensure constant 
row sum we choose the inter-layer coupling as $\sigma_{\text{inter}}=\sigma^{12}=\sigma^{32}$, $\sigma^{21}=\sigma^{23}=\sigma^{32}/2$ and 
$\sigma^{11}=\sigma^{13}=\sigma^{22}=\sigma^{31}=\sigma^{33}=0$.
The inter-layer interaction is only through the activator variables, i.e., the coupling matrix $\mathbf{G}$, whereas the intra-layer interaction is realized through rotational coupling with coupling matrix $\mathbf{H}$
\begin{equation}
\matr{G}=\m{\varepsilon^{-1}& 0\\ 0 & 0},\,\matr{H}=\m{\varepsilon^{-1}\cos \phi& \varepsilon^{-1} \sin \phi\\ -\sin \phi&\cos \phi}
\end{equation}
and coupling phase $\phi = \frac{\pi}{2}-0.1$. The latter coupling scheme, which consists predominantly of activator-inhibitor cross-coupling, is similar to a phase-lag of approximately $\pi/2$ in the Kuramoto phase oscillator model and has been chosen such that chimera states are most likely to occur~\cite{OME13}. Chimera states exhibit typical arc-shaped mean 
phase velocity profiles. 

Note that the hub receives input from $N$ nodes in each outer layer, while each node of the outer layers receives inter-layer input only from the hub. In order to balance the inputs, and secure that the input which the hub receives does not outweigh the local dynamics by orders of magnitude, 
we introduce the normalization factor $\frac{1}{N}$ in the coupling term in Eq.(\ref{eq:my_equation_relay}), which scales the coupling term such that it becomes of the same order as the local hub dynamics; thus we have an active relay. If we drop the normalization factor $\frac{1}{N}$, which 
we will do only in Sect.\,III, the local hub dynamics does not play a role 
any more, and the hub becomes effectively a passive relay. 

To detect partial relay synchronization patterns between the outer layers, we compute the local synchronization error $E_i^{13}$:

\begin{equation}
    E_i^{13} = \lim_{T\to\infty}\frac{1}{T}\int_0^T \left \Vert \vec{x}_i^1(t)-\vec{x}_i^3(t) \right \Vert dt.
\end{equation}

This measure calculates the difference in variables $\mathbf{x}$ between two corresponding nodes with index $i$ in the outer layers, averaged over 
time $T$~\cite{SAW18c}. Since we calculate the errors only between pairs of nodes, this measure is well suited to detect partial relay synchronization patterns, where only parts of the outer layers are synchronized with 
each other.

Introducing a time delay $\tau$ in the coupling term often leads to travelling patterns as shown, e.g., in Ref. \onlinecite{SAW17}. To avoid any numerical artifacts caused by those, we will make use of the method of detrending introduced in Ref. \onlinecite{SAW18c}. By detrending the data we 
can avoid this problem: After each time-step in the numerical simulation we re-index the nodes $i$ in such a way that $i' = (i + c)|_N$, where $|$ stands for the absolute value and $c$ is given by the center of the largest domain of the ring where for all $i$'s of that domain $\left \Vert \vec{x}_i^m(t) - \vec{x}_{i+1}^m(t) \right \Vert < \theta$. As the threshold parameter $\theta$ we choose $\theta = 0.25$.

\section{Partial synchronization of chimera states}

\begin{figure}
\includegraphics[width=1.0\linewidth]{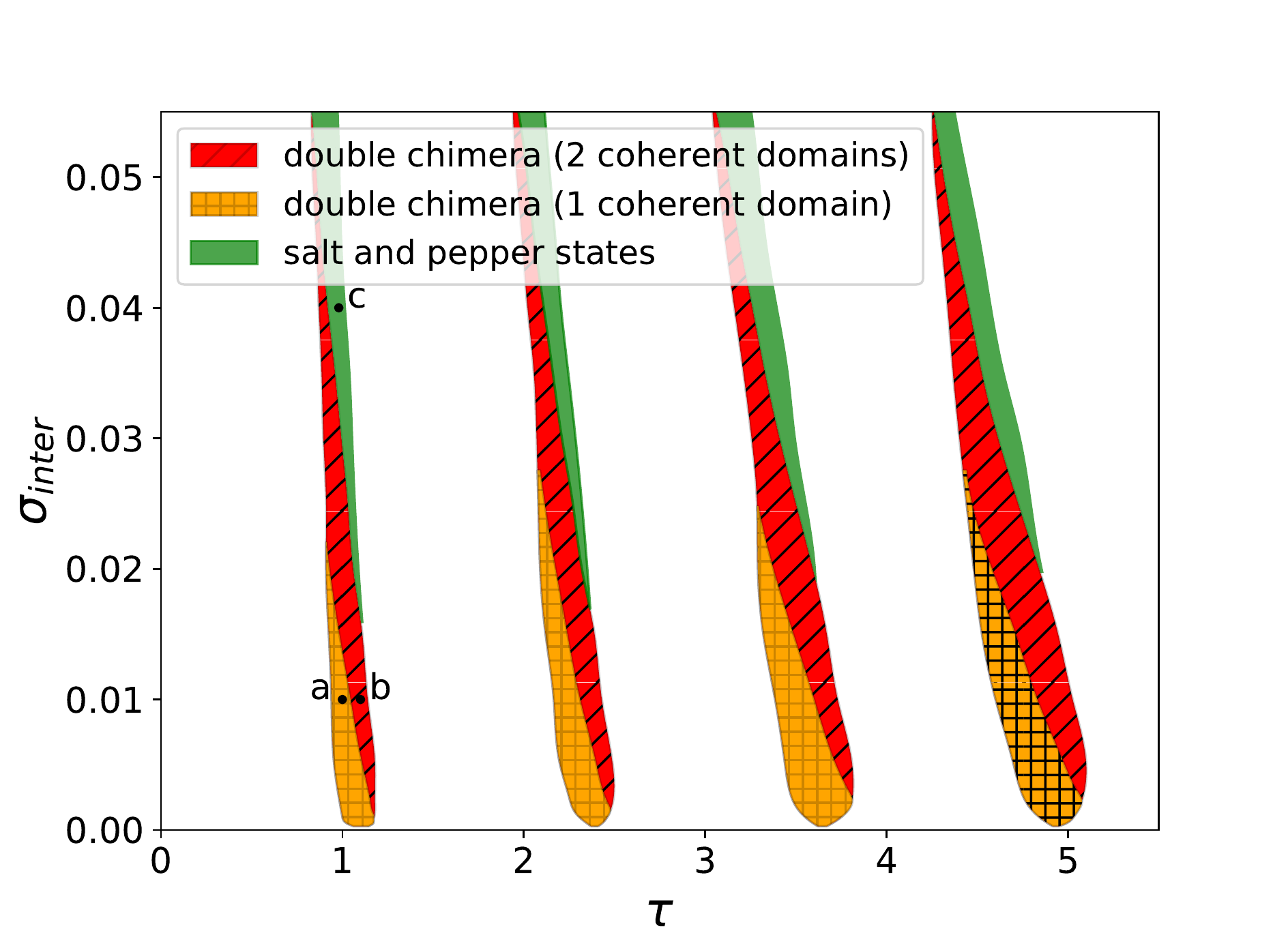}
\caption{Partial synchronization tongues in the parameter plane of the inter-layer coupling strength $\mathbf{\sigma_{\text{inter}}}=\sigma^{12}$ and time delay $\mathbf{\tau}$. Partial synchronization patterns including chimera states (cross-hatched orange), double chimera states (hatched 
red), and salt and pepper states (green) emerge in tongue-like regions. All results have been obtained with Eqs.\,(\ref{eq:my_equation_outer}) and 
(\ref{eq:my_equation_relay}), where Eq.\,(\ref{eq:my_equation_relay}) has 
been used without the normalization factor $\frac{1}{N}$ in the coupling term (passive relay). The parameters are $N=500$, $R^m=170$, $\sigma_{\text{intra}}=0.2$, $\epsilon = 0.05$, $a=0.5$.}
\label{fig:snapshots_Nagies}
\end{figure}

To analyze the spatiotemporal dynamics of our three-layer network, we run 
numerical simulations for random initial conditions with varying inter-layer time delay $\tau$ and inter-layer coupling strength $\sigma_{\text{inter}}$. Since we are particularly interested in synchronization patterns between chimera states in the outer layers, we set the parameters controlling intra-layer coupling and topology to values which are known to induce chimeras in isolated layers~\cite{OME13}. Chimera states are characterized by the spontaneous emergence of coexisting coherent and incoherent domains on completely symmetrical networks. Especially intriguing is their resemblance with some patterns observed in brain dynamics like epileptic seizures~\cite{GER20}. 

\begin{figure}
\includegraphics[width=1.00\linewidth]{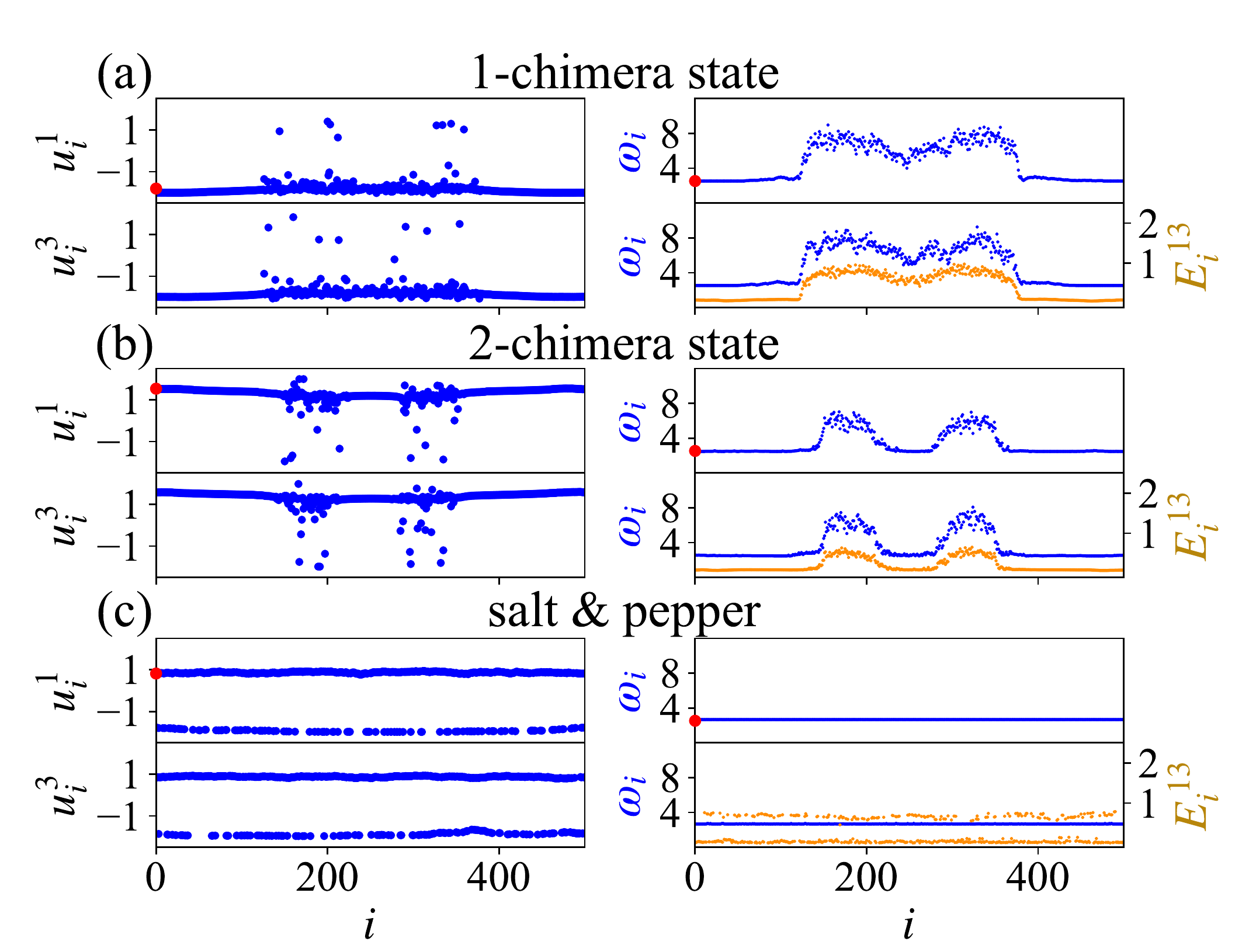}
\caption{Dynamics of the outer layers for different values of the inter-layer coupling strength $\mathbf{\sigma_{\text{inter}}}=\sigma^{12}$ and 
time delay $\mathbf{\tau}$, marked in Fig.\,\ref{fig:snapshots_Nagies}: \textbf{(a)} double chimera state with 1 coherent domain for $\sigma_{\text{inter}}=0.01$, $\tau=1.0$ \textbf{(b)} double chimera state with 2 coherent domains for $\sigma_{\text{inter}}=0.01$, $\tau=1.1$ \textbf{(c)} salt and pepper state for $\sigma_{\text{inter}}=0.04$, $\tau=0.98$. The left column shows snapshots of variables $u_i^m$ for the outer layers $m=1,3$ (dark blue), whereas the right column shows the corresponding mean phase velocity profiles $\omega_i$ (dark blue) for each layer and local inter-layer synchronization error $E^{13}_i$ (light orange), averaged over a time window $t_{\textrm{a}} = 7\,000$. The snapshot of $u^2$ and the mean phase velocity of the hub are marked by red dots in the upper panels. In panels (a) and (b), the coherent parts of the chimera states in the outer layers are synchronized, whereas the incoherent parts 
are desynchronized (double chimera states). Other parameters as in Fig.\,\ref{fig:snapshots_Nagies}.}
\label{fig:dynamic_Nagies}
\end{figure}

The results of our numerical simulations are summarized in Fig.\,\ref{fig:snapshots_Nagies}, which shows the parameter plane of inter-layer coupling strength and inter-layer time delay with the domains of different observed partial relay synchronization patterns, organized in evenly spaced tongue-like regions. At least two more tongues exist for larger time delays but are not mapped out in detail. The cross-hatched orange and hatched red domains are so-called double chimeras~\cite{SAW18c}. The name refers to the twofold occurrence of the typical chimera-like feature of the coexistence of coherence and decoherence: On the one hand within each outer layer, where we observe regular chimera states with a single or multiple coherent domain, and on the other hand in the partial inter-layer synchronization of chimera states (only the coherent cores of both outer layers are synchronized). The green areas in the synchronization tongues denote {\em salt and pepper states}, which are characterized by strong variations 
on very short length scales, so that the dynamical patterns have arbitrarily short wavelengths~\cite{KON10,BAC14,SEM17,SAW19}.

The different partial synchronization patterns are shown in more detail in Fig.\,\ref{fig:dynamic_Nagies} for three sets of parameters marked by black dots and labeled a,b,c in Fig.\,\ref{fig:snapshots_Nagies}: In the cross-hatched orange region we observe double chimeras comprising regular chimera states with one coherent domain in both outer layers (Fig.\,\ref{fig:dynamic_Nagies}(a)), while the hatched red region exhibits double chimeras with two coherent domains in both outer layers (Fig.\,\ref{fig:dynamic_Nagies}(b)). This is clearly shown by the snapshots (left column) and 
the mean phase velocity profiles (right column). The local synchronization error (right column, orange) indicates partial relay synchronization of 
the coherent domains. Note that the hub (red dot) synchronizes with the coherent domain. Ultimately we can also observe partially synchronized salt and pepper states in the outer layers for higher inter-layer coupling strengths, marked by the green areas (Fig.\,\ref{fig:dynamic_Nagies}(c)). In salt and pepper states all oscillators move with the same mean phase velocity. In contrast to the completely synchronized state a large number of oscillators split from the main coherent cluster and move with a constant phase lag (on a slightly perturbed limit cycle)~\cite{SAW19,SAW19a}. 

We can observe a resonance effect with respect to the time delay $\tau$: The partial relay synchronization tongues appear close to full or half integer multiples of the intrinsic period of the uncoupled system $T=2.3$ 
(see Fig.\,\ref{fig:snapshots_Nagies}). Furthermore we can see that the tongues are tilted to the left with increasing inter-layer coupling strength $\sigma_{\text{inter}}$. In Ref. \onlinecite{SAW18c} analytical calculations were performed for a triplex network~\footnote{Note that in the present work the relay-layer of this triplex network is reduced to a single 
node, but the results carry over.} yielding a decreasing period $T$ for increasing inter-layer coupling strength, which is consistent with the observed tilt of the tongues in Fig.\,\ref{fig:snapshots_Nagies}.

In between the tongues of Fig.\,\ref{fig:snapshots_Nagies} and for the special case of no time delay ($\tau = 0$) we observe desynchronization or complete synchronization or traveling waves where all nodes move with the same mean phase velocity but with a constant phase lag between them. For high inter-layer coupling strengths the tongues are only shown up to $\sigma_{\text{inter}}=0.055$. Salt and pepper states cease to exist at approximately $\sigma_{\text{inter}}=0.06$, and only completely synchronized states are observed. The simulations with the lowest inter-layer coupling strength for which partial relay synchronization still occurs are done with $\sigma_{\text{inter}} = 0.0002$. Below that value only desynchronization has been observed. The transition from desynchronization to chimera states is in the focus of the next section.

\section{Induction of chimera states via the hub}

\begin{figure}
\includegraphics[width=1.0\linewidth]{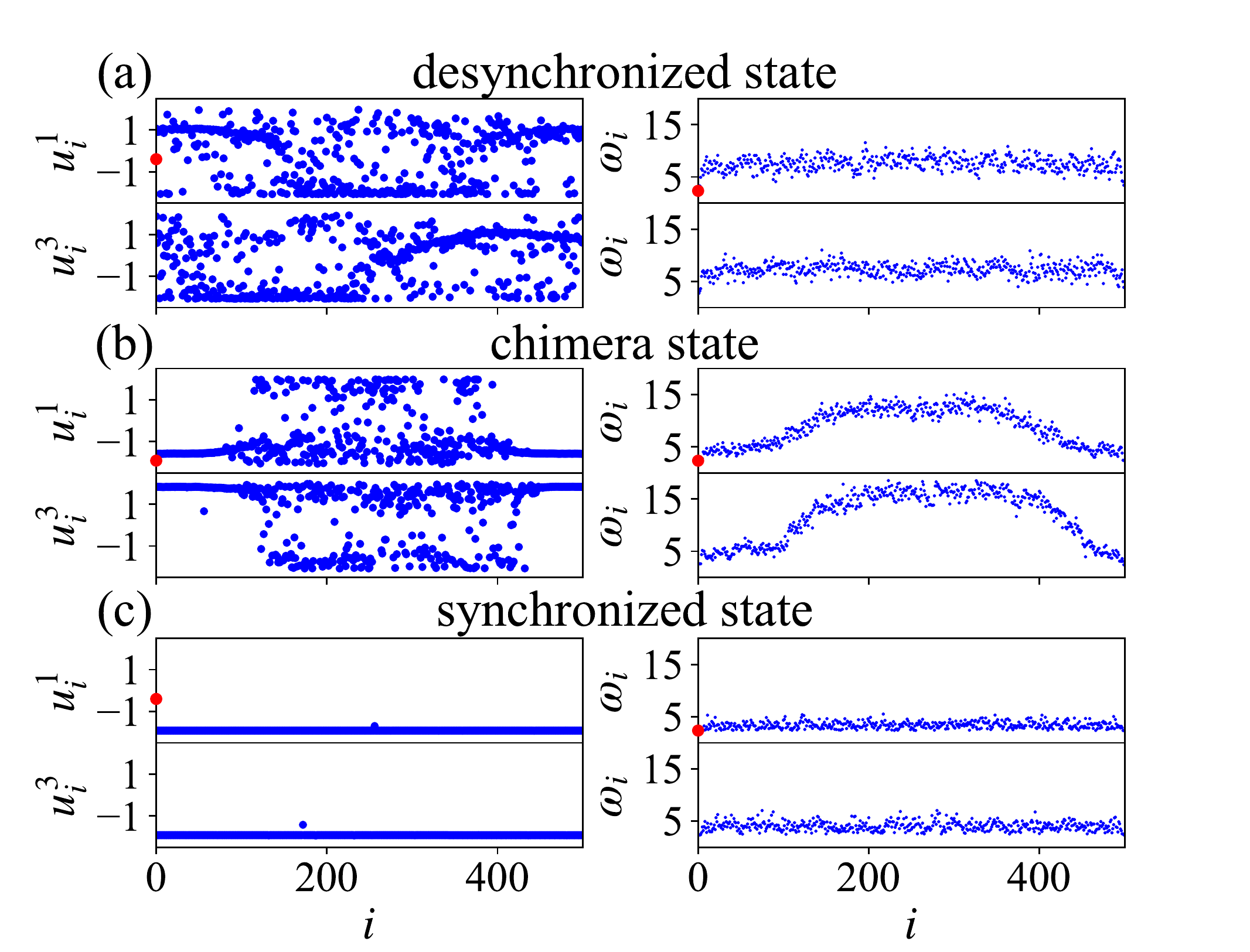}
\caption{Chimera states induced via a hub: dynamics for different values of weak inter-layer coupling strength $\sigma_{\text{inter}}$ and time delay $\tau$, marked by black circles (a,b,c) in Fig.\,\ref{fig:heatmap_combined}. \textbf{(a)} desynchronized state for $\sigma_{\text{inter}}=0$, $\tau=0$ \textbf{(b)} chimera state in both layers for $\sigma_{\text{inter}}=0.00006$, $\tau=1.08$ \textbf{(c)} completely synchronized state for $\sigma_{\text{inter}}=0.00015$, $\tau=1.14$. Left panels: Snapshots of the activator variable $u_i^m$ of the outer layers $m=1$ and $m=3$. The snapshots have been taken for the last time step of the simulation. Right panels: Mean phase velocity profiles $\omega_i$ of both outer layers. The snapshot of $u^2$ and the mean phase velocity of the hub 
are marked by red dots in the upper panels. The mean phase velocity has been averaged over the time window $t_{\textrm{a}} = t_{\textrm{s}} - t_{\textrm{trans}} = 8\,000$ with the total simulation time $t_{\textrm{s}}=10\,000$ and the transient time $t_{\textrm{trans}}=2\,000$. $\sigma_{\text{intra}}=0.00055$, other parameters as in Fig.\,\ref{fig:snapshots_Nagies}, but with the normalization factor $\frac{1}{N}$ in the coupling term (active relay).}
\label{fig:chaotic_state}
\end{figure}

We will now examine whether the hub can induce chimera states through inter-layer coupling in the outer layers if the single individual outer layers exhibit no chimera states. Accordingly, we investigate if the interplay of intra-layer coupling, inter-layer coupling, and time delay can result in chimera states which are not observed in the isolated outer layers. The regime of chimera states has been explored in the parameter plane of the intra-layer coupling strength $\sigma_{\text{intra}}$ and the coupling radius $r$ in Ref. \onlinecite{OME13}. With no connection to the hub ($\sigma_{\text{inter}}=0$), the value $\sigma_{\text{intra}}=0.00055$ is determined for which there are no chimera states in the outer layers after a simulation time of $t_{\textrm{s}}=10\,000$. For larger coupling 
strength $\sigma_{\text{intra}}>0.00055$, the dynamics of the isolated outer layers shows chimera states. Figure \ref{fig:chaotic_state}(a) shows the dynamics of the system at an intra-layer coupling value of $\sigma_{\text{intra}}=0.00055$ without inter-layer coupling to the hub. 

Throughout the rest of this paper we will use the normalization factor $\frac{1}{N}$ in the coupling term in Eq.(\ref{eq:my_equation_relay}), such 
that the hub becomes an active relay, in contrast to the passive relay which we have considered in Figs.\,\ref{fig:snapshots_Nagies} and \ref{fig:dynamic_Nagies}. This is appropriate since we are now considering very small values of the intra-layer and inter-layer coupling strengths, and there should be some balance between the local dynamics and the coupling terms.  

Figure \ref{fig:chaotic_state} shows the transition from a desynchronized 
state, when the hub is disconnected from the outer layers (Fig.\,\ref{fig:chaotic_state}(a)), to a completely synchronized state, when the inter-layer coupling strength is increased (Fig.\,\ref{fig:chaotic_state}(c)). If we choose appropriate values for the inter-layer coupling strength and time delay, chimera states can be induced in the outer layers by the hub (Fig.\,\ref{fig:chaotic_state}(b)). Furthermore, we investigate the hub dynamics. In the case where chimera states appear in both outer layers, snapshots and space-time plots reveal that the hub is not phase-synchronized with the coherent cores of the chimera states in the outer layers. On the contrary, the space-time plots (not shown here) demonstrate that the hub dynamics has a phase lag to that of the coherent core of the chimera states. 

\begin{figure}
\includegraphics[width=1.0\linewidth]{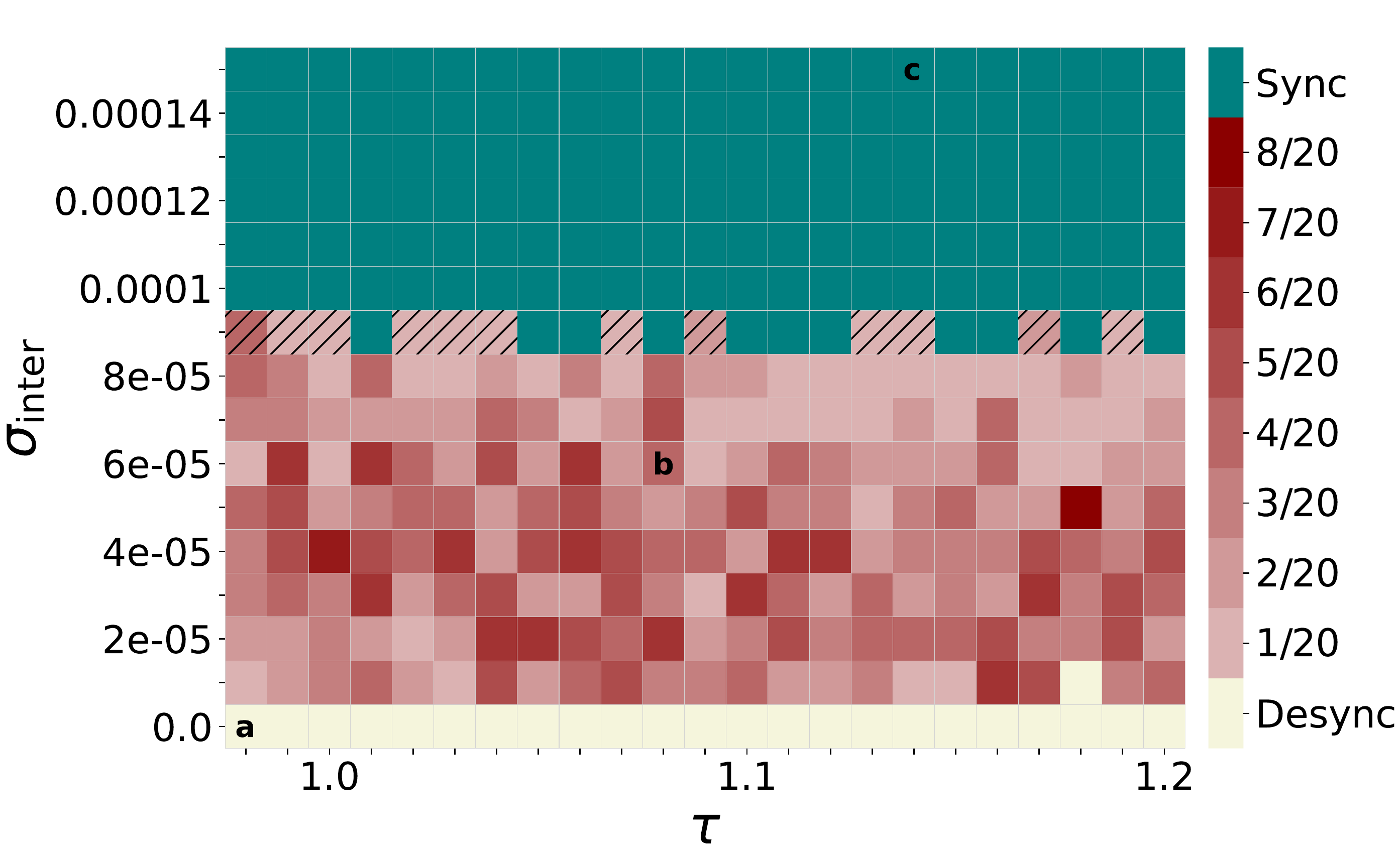}
\caption{Chimera states in the parameter plane of inter-layer coupling strength $\mathbf{\sigma_{\mathit{\text{inter}}}}$ and time delay $\mathbf{\tau}$. The color of each tile indicates the state for specific values for $\sigma_{\text{inter}}$ and $\tau$ for 20 simulations with random initial conditions, respectively. Light yellow tiles represent desynchronized states, while turquoise tiles stand for states which are almost completely synchronized. All red tiles indicate chimera states which emerge in both layers. The intensity of the red color indicates the ratio of simulations in which chimera states emerge. In the hatched region synchronization can be found additionally. In the other cases desynchronization can be observed in at least one layer. The intra-layer coupling strength within both outer layers is set to $\sigma_{\text{intra}}=0.00055$ for all simulations. The total simulation time is set to $t_{\textrm{s}}=10\,000$ for all simulations. The dynamics of the three marked dots are shown in Fig.\,\ref{fig:chaotic_state}(a), (b), (c). Other parameters as in Fig.\,\ref{fig:chaotic_state}.}
\label{fig:heatmap_combined}
\end{figure}

Figure \ref{fig:heatmap_combined} summarizes the results in the parameter 
plane of inter-layer coupling strength $\sigma_{\text{inter}}$ and time delay $\tau$, where 20 simulations with random initial conditions are performed for each set of parameters. The figure indicates that without coupling of the hub to the outer layers ($m=1, 3$), no chimera states occur (light yellow). For large $\sigma_{\text{inter}}$ both layers are fully synchronized through the hub (turquoise). By varying $\sigma_{\text{inter}}$ and $\tau$, chimera states can be induced for some parameter combinations of $\sigma_{\text{inter}}$ and $\tau$. We have chosen $\tau \simeq 1$ 
because the first partial relay synchronization tongue in Fig.\,\ref{fig:snapshots_Nagies} is located at this value. If the simulations are repeated several times using different random initial conditions, different patterns may be found because of the high multistability of the system.
When a chimera state emerges in one outer layer, in most cases no corresponding chimera state can be observed in the other outer layer. In some cases, however, the chimera states occur in both layers, and then we represent the number of such simulations by the red color intensity. Note that the chimera states which appear in both layers do not exhibit full or partial relay synchronization. This can be attributed to the low intra-layer 
coupling strength at with chimera states in the outer layers are just beginning to emerge. 

In conclusion, we have shown that coupling the initially isolated and desynchronized outer layers via the hub can induce chimera states in these layers. We have, however, investigated a parameter regime of $\sigma_{\text{inter}}$ and $\tau$ where chimera states are just beginning to emerge. Therefore the mean phase velocity profiles are not as pronounced as for higher values of intra-layer coupling strength.

\section{Dilution of links with the hub}

In the following, we investigate to what extent a dilution of the inter-layer connections between the outer layers and the hub plays a role for the chimera states in the outer layers. In particular, we address the question of whether the chimera states will persist if the links either to their coherent domain or to their incoherent domain are removed. Consequently, the inter-layer links between the hub and the outer layers are removed 
in different percentages of the links to coherent or incoherent nodes. The connections are cut in a controlled regular manner. For example, for a dilution $d=50 \, \%$ of the links with the coherent core, each second connection to the coherent core is removed. The effects of dilution of network links upon synchronization has also been studied in power grid models~\cite{TUM18,TUM19}.

\begin{figure}
\includegraphics[width=.93\linewidth]{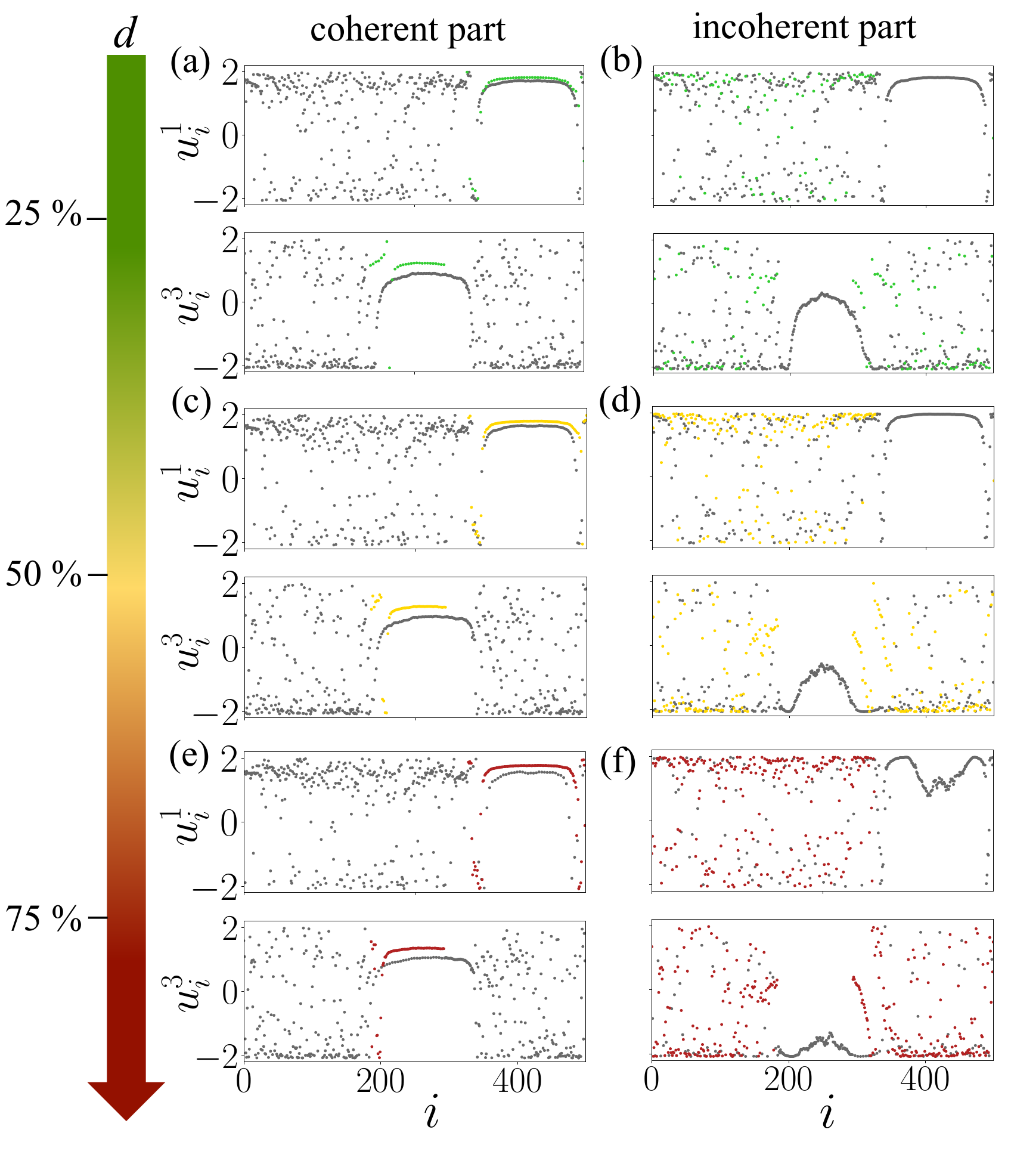}
\caption{Effect of dilution on chimera states in the outer layers: snapshots for three values of the dilution percentage (a,b) $d=25 \, \%$, (c,d) $d=50 \, \%$, (e,f) $d=75 \, \%$. Left column: $d$ links of the coherent domain are cut, right column: $d$ links of the incoherent domain are cut. The color-coded dots correspond to the $u$-value of the nodes in the outer layers to which the link from the hub was cut, whereas gray dots are still connected to the hub. Green dots: $d=25\, \%$ links cut (a) 
of the coherent domain (70 links) or (b) of the incoherent domain (178 links). Yellow dots: $d=50 \, \%$ links cut (c) of the coherent domain (140 links) or (d) of the incoherent domain (358 links). Red dots: $d=75 \, \%$ links cut (e) of the coherent domain (210 links) or (f) of the incoherent domain (538 links). Parameters: $\tau=1.08$, $\sigma_{\text{inter}}=0.00006$, $\sigma_{\text{intra}}=0.00055$, simulation time $t_{\textrm{s}}=25\,000$. Other parameters as in Fig.\,\ref{fig:chaotic_state}.}
\label{fig:dilution_verlauf_snapshots}
\end{figure}

\begin{figure}
\includegraphics[width=.9\linewidth]{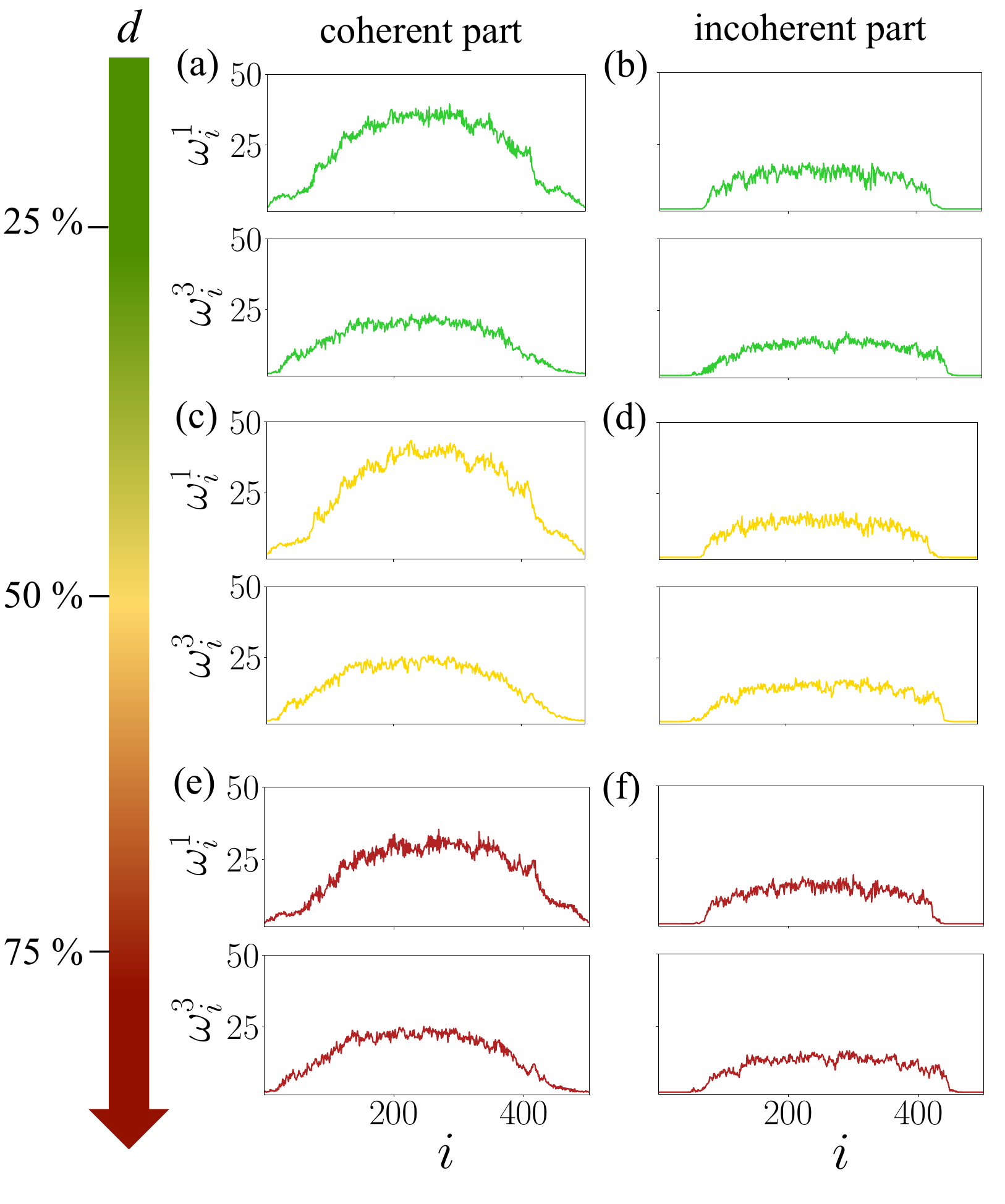}
\caption{Effect of dilution on chimera states in the outer layers: mean phase velocity profiles averaged over $t_{\textrm{a}}=10\,000$ for the same parameters as in Fig.\,\ref{fig:dilution_verlauf_snapshots}. }
\label{fig:dilution_verlauf_omega}
\end{figure}

We start the dilution at time $t=15\,000$, where $t$ corresponds to the 
time at which we have determined the dynamics of the chimera states occurring in both outer layers in Fig.\,\ref{fig:heatmap_combined}. To study the dilution effects upon the system, the simulation is run until $t_{\textrm{s}}=25\,000$. As a measure of dilution, we introduce the percentage 
$d$ of removed links. It represents the percentage of cut connections between the hub ($m=2$) and the coherent or incoherent core of the chimera 
states in the layers $m=1, 3$. The effects of dilution for $d=25 \, \%, \, 50 \, \% \; \textrm{and} \; 75 \, \%$ are demonstrated in Figs.\,\ref{fig:dilution_verlauf_snapshots} and \ref{fig:dilution_verlauf_omega}.

A dilution of the inter-layer links to the incoherent core of the chimera 
states at different percentages does not cause a visible deterioration or 
even destruction of the chimera states. The corresponding snapshots in Fig.\,\ref{fig:dilution_verlauf_snapshots}(b), (d), (f) are indicating the emergence of chimera states. The mean phase velocity profiles confirm this phenomenon in Fig.\,\ref{fig:dilution_verlauf_omega}(b), (d), (f), where all profiles show the typical arc-shape with a coherent and an incoherent part of a chimera state regardless of dilution to the incoherent region.

The dilution of links to the coherent core of the chimera states leads to 
splitting-off some nodes from the original coherent domain in layer 3. In 
Fig.\,\ref{fig:dilution_verlauf_snapshots}(a), (c), (e), the nodes of the 
coherent domain to which the connection has been broken oscillate with a phase different from that of the original coherent core nodes. The splitting of the coherent core yields a modified detrended mean phase velocity profile exhibiting higher values in case of dilution, as shown in Fig.\,\ref{fig:dilution_verlauf_omega}(a), (c), (e). 
The increased values in the incoherent region can be attributed to the increased number of split-off nodes resulting from the dilution. On the other hand, the formation of another coherent range of $u$-values smears out 
the coherent region in the mean phase velocity profile which can no longer be clearly distinguished. The fact that a significant difference exists 
between the two coherent cores implies that the mean phase velocities are 
increased, and the constant regions vanish. For $d=50 \, \%$ dilution, this behavior is more pronounced in the first layer than in the third layer, which can be attributed to the different initial conditions underlying the two chimera states.  
coherent core as demonstrated in Fig.\,\ref{fig:dilution_verlauf_omega}(e). 
The increase of the mean phase velocity profile with detrending in the upper panel of Fig.\,\ref{fig:dilution_verlauf_omega}(c) can be traced back 
to the division of the coherent cores' velocities.

We can therefore conclude that the dilution of links to the coherent part 
of the chimera states has the most significant effect at $d=50 \, \%$ since here the size of the newly formed coherent core of each layer is the 
largest. This assumption has been confirmed by simulations in which the dilution of the inter-layer links has been applied at an earlier stage of the simulation when chimera states have not yet formed. Our investigation 
shows that at this stage diluting inter-layer links to the coherent core has a much more pronounced negative effect upon the formation of chimera states than dilution of inter-layer links to the incoherent core.

\section{\label{sec:Conclusion}Conclusion}

The focus of this paper is the effect of inter-layer coupling delay and dilution upon chimera states in three-layer networks with a hub. While previous works have analyzed triplex networks of three identical layers~\cite{SAW18c,SAW18,SAW19a}, here we have chosen the (middle) relay layer as a 
hub, i.e., a single node. This is typical of the brain architecture, where distant brain areas communicate via a functional relay (e.g., thalamus or hippocampus) consisting of a significantly smaller number of neurons. First, we have elaborated the role of the hub and its importance for the existence of chimera states. We have mapped out the regimes of partial relay synchronization of chimera states between two remote layers in the parameter space of inter-layer coupling strength and time delay, and found tongue-like regions in the parameter plane exhibiting double chimeras, i.e., chimera states in the remote layers whose coherent cores are synchronized with each other, either with one or with two coherent cores, and salt-and-pepper states consisting of two intertwined groups of oscillators, all moving with the same phase velocity.

Secondly, we have considered a very low intra-layer coupling strength such that no chimera states occur in the outer layers when the inter-layer coupling to the hub is switched off. For certain combinations of $\tau$ and $\sigma_{\text{inter}}$ we have found chimera states induced by the hub. In more detail, chimera states occur in one remote layer only as well 
as in both layers, for different random initial conditions, because of the high multistability of the system.

In the third part, we have investigated the influence of dilution of links to the hub upon the dynamics. Starting with chimera states induced by the hub, we have cut the inter-layer links either to the coherent domain or to the incoherent domain of the chimera states at a time when the chimera states have already emerged. The greatest effect of dilution has been observed when $d=50 \, \%$ of the links to the coherent domain were removed. In contrast, no significant change in the chimera states could be found when links to the incoherent domain were removed. 

These results on partial relay synchronization scenarios induced by a passive or active hub might be useful for a deeper understanding of dynamical patterns and functions occurring in the brain. 

\begin{acknowledgments}
This work was supported by the Deutsche Forschungsgemeinschaft (DFG, German Research Foundation, project No. 429685422). We thank J\"urgen Kurths for his hospitality at the Potsdam Institute for Climate Impact Research and Sebastian Nagies for fruitful discussions.
\end{acknowledgments}

\section*{Data Availability Statement}
The simulation data that support the findings of this study are available 
within the article. 

\nocite{*}
\bibliographystyle{apsrev4-1}
\bibliography{SAW21_final.bbl}


\end{document}